\documentstyle[preprint,aps]{revtex}
\begin{document}

%\baselineskip=18pt plus 2pt minus 1pt
%\magnification=1200
%\hsize=5.7truein
%\vsize=8.4truein
%\voffset=24pt
%\hoffset=.1in
%
% Title Page
%

\title{ Antiferromagnetic and van Hove Scenarios
for the Cuprates: Taking the Best of Both Worlds.}

\author{ Elbio Dagotto, Alexander Nazarenko and Adriana Moreo}

\address{Department of Physics and National High Magnetic Field Lab,
Florida State University, Tallahassee, FL 32306, USA}

\date{\today}
\maketitle

\begin{abstract}

A theory for the high temperature superconductors is proposed.
Holes are spin-1/2, charge e, quasiparticles strongly
dressed by spin fluctuations. Based on their dispersion,
it is claimed that the experimentally observed van Hove singularities
of the cuprates are likely
originated by antiferromagnetic (AF) correlations. From the
two carriers problem in the 2D t-J model,
an effective Hamiltonian for holes is defined with
no free parameters.
This effective model has superconductivity in the
${\rm d_{x^2-y^2}}$ channel, a critical temperature ${\rm T_c \sim 100K}$
at the optimal hole density, ${\rm x=0.15}$, and a quasiparticle lifetime
linearly dependent with energy.
Other experimental results are also $quantitatively$ reproduced by
the theory.

\end{abstract}


%
% Introduction
%

Recent experimental results for the high temperature cuprate
superconductors have suggested
that the pairing state is highly anisotropic, probably a ${\rm d_{x^2-y^2}}$
singlet. In theories where the pairing mechanism is
produced by antiferromagnetic fluctuations,
superconductivity in the d-wave channel appears
naturally as shown in the context of the 2D Hubbard model
using self-consistent techniques,\cite{bickers} phenomenologically with
the nearly-antiferromagnetic Fermi liquid state,\cite{pines} and
using various techniques in the 2D t-J model.\cite{tjpapers} While these
approaches seem
successful in the prediction of the superconducting
symmetry, some
phenomenological details of the cuprates remain hidden like the
existence of an ``optimal'' doping.
A different family of theories for the cuprates makes extensive
use of the concept of van Hove (vH) singularities in the
 density of states (DOS).\cite{tsuei}
In this context, the quasiparticle dispersion is
sometimes extracted from angle-resolved photoemission (ARPES)
experiments or band structure calculations,
a vertex interaction is proposed (usually invoking electron-phonon or
excitonic mechanisms), and predictions for superconductivity are
made using standard techniques.
The strong point of this approach is the natural existence of
an optimal doping which occurs when the chemical potential reaches the vH
singularity.

The purpose of this paper is to describe a microscopically-based
theory of the cuprates that combines the strong features of
the above described antiferromagnetic and van Hove scenarios.
The first step in the construction of such a theory
is the observation that the ARPES quasiparticle dispersion
may be caused by holes moving in a local antiferromagnetic
environment, rather than by band structure effects.
This idea is motivated by the existence of universal
flat bands\cite{gofron,dessau,ma}
near ${\bf k} = (\pi,0),(0,\pi)$
in the spectrum of hole-doped cuprates (2D square lattice language), which are
difficult
to understand unless caused by correlation
effects in
the  ${\rm Cu O_2}$ planes.
Recently,\cite{flat-theory,flat2} it has been shown
that models of correlated electrons can account for such flat bands,
and here we further elaborate on this idea showing that
the agreement with ARPES is $quantitative$.

Using the well-known two dimensional (2D)
${\rm t-J}$ model defined by the Hamiltonian,
\begin{equation}
{\rm H_{tJ}=-t\sum_{<i,j>\sigma} (c^+_{i\sigma} c_{j\sigma}+h.c.)
  +J\sum_{<i,j>}({\bf S}_i\cdot{\bf S}_j-{1\over4} n_in_j)  },
\end{equation}
in the standard notation, the dispersion of one hole in an
antiferromagnet can be calculated accurately\cite{flat-theory} with
numerical or analytical techniques.
At small J/t, it was found that the hole dispersion is
\begin{equation}
\epsilon_{\bf k} = {\rm 1.33 J cosk_x cosk_y + 0.37 J (cos2k_x + cos2k_y)},
\end{equation}
which was calculated using a Green's Function
Monte Carlo method.\cite{flat-theory}
${\rm J=0.125 eV}$ is the actual scale of the problem, and Eq.(2) shows that
holes move within the
same sublattice to avoid distorting the AF background.
To improve the quantitative agreement with experiments described below,
here a small hopping amplitude along the plaquette diagonals
${\rm t' = 0.05t}$ has been included in the
Hamiltonian to produce the
dispersion Eq.(2), but the qualitative
physics presented in this paper is the same as long as ${\rm |t'/t|}$ is
small. Now, let us
assume a rigid band picture for the quasiparticles.\cite{flat-theory}
The dispersion is plotted in Fig.1a against momentum with the Fermi level
$at$ the flat band which corresponds to a hole density of ${\rm x=0.15}$ (in
Fig.1a and 1c below, $\epsilon_{\bf k}$ is inverted i.e.
the electron language is
used rather than the hole language to facilitate the comparison with
experiments).
$\epsilon_{\bf k}$ contains
a saddle-point located close to ${\bf k} = (\pi,0),(0,\pi)$,
which induces a large DOS in the spectrum.
In addition, $\epsilon_{\bf k}$ is nearly degenerate along
the ${\rm cosk_x + cosk_y = 0}$ line increasing the DOS
in the vicinity of the Fermi level of Fig.1a.
All these qualitative features are common to several models of
correlated
electrons, and should not be considered as exclusively produced by
the t-J model.

What is the influence of a finite hole density on this dispersion? Recent
experiments by Aebi et al.\cite{aebi} have shown the presence of strong
antiferromagnetic (AF) correlations in the normal state of BSSCO leading to the
formation of hole pockets in their results.
A rigid-band filling of $\epsilon_{\bf k}$
reproduces qualitatively their data (Fig.1b).
The most important
detail to consider at a small but finite hole density is that
the quasiparticle weight is smaller for the bands centered at
momentum $(\pi,\pi)$ than those at $(0,0)$
(as represented pictorially in Fig.1a-b, with dashed lines).
These are the ``shadow bands'' which were discussed before
in the spin-bag approach.\cite{spinbag}
Another argument
in favor of using our hole dispersion at a finite density also comes from
experiments. In Fig.1c we
compare $\epsilon_{\bf k}$ along the ${\rm {\bf k} = (0,0)
- (\pi,0) }$
direction against ARPES results  by the Argonne
group\cite{gofron,campuzano} for YBCO. The agreement is excellent.
It is worth emphasizing
that the theoretical curve of Fig.1c is derived from a microscopic
Hamiltonian, and it is $not$ a fit of ARPES data. This is a major
difference
between the present approach and previous vH scenario calculations.

To further elaborate
on whether our dispersion ( derived
at ${\rm x \rightarrow 0}$) can be used at finite ${\rm x}$,
we have carried out extensive
QMC numerical simulations of the 2D Hubbard model at several
densities, a large coupling ${\rm U/t=8}$ (i.e. in the regime of the t-J
model),
and temperature ${\rm T=t/4}$, where
${\rm t \approx 0.4eV}$. The results (Fig.2a) show
that nearest and next-to-nearest neighbor spins tend to be antiparallel
even with hole densities as large as ${\rm x=0.25}$.\cite{duffy}
Reducing the temperature, the correlations would be even stronger
than those in Fig.2a. Thus, the assumption
that the dispersion Eq.(2) holds near half-filling
is supported by numerical and experimental
evidence.\cite{trugman} The intuitive picture to
remember is that as long as the antiferromagnetic correlation length
is larger than the typical size of a quasiparticle, they will
behave as if moving in a nearly perfect AF background.

As a second step in building up a model for the cuprates,
the interaction among the hole quasiparticles is necessary, and
it will be constructed based again
on results obtained for the 2D t-J model. In this case,
it is well-known that an
effective attractive force exists in an antiferromagnet leading to the bound
state of two holes in the d-wave channel.\cite{review}
Since this problem is non-trivial, we
will simplify its analysis by studying the potential in the atomic limit
(large J/t) where
the attraction is induced by the minimization of the number of
missing antiferromagnetic links\cite{review} (similar
to the attraction among spin-bags in a spin-density-wave
background\cite{spinbag}).
Assuming that the link spin-spin correlation is not much distorted by the
carriers in this limit, the binding energy of two holes is
$\Delta_B = e_{2h} - 2 e_{1h} = {\rm J ( \langle {\bf S_{\bf i}}\cdot{\bf
S_{\bf j}} \rangle - 0.25)}$,
where $e_{nh}$ is the energy of $n$ holes with respect to the
antiferromagnetic ground state energy, and ${\bf i}$ and ${\bf j}$ are nearest
neighbors. Accurate numerical
simulations\cite{carlson} have shown that $\langle {\bf S_{\bf i}}\cdot{\bf
S_{\bf j}} \rangle
\approx -0.3346$, and thus $\Delta_B \approx {\rm -0.6 J}$. To  mimic this
effect, an attractive term is introduced
that reduces the energy when two quasiparticles share the same
link. Thus, the Hamiltonian proposed here is
\begin{equation}
{\rm H =-\sum_{{\bf k},\alpha} \epsilon_{\bf k}
c^{\dagger}_{{\bf k}\alpha} c_{{\bf k}\alpha}
- |V| \sum_{\langle {\bf ij} \rangle} n_{\bf i} n_{\bf j} },
\end{equation}
where ${\rm c_{{\bf k}\alpha}}$ is an operator that destroys a
quasiparticle with momentum ${\rm {\bf k}}$ and in sublattice
$\alpha = {\rm A,B}$;
${\rm n_{\bf i}}$ is the number operator at site ${\bf i}$; ${\rm |V|
= 0.6J}$, and $\epsilon_{\bf k}$ is given in Eq.(2)\cite{comm}. Since in
the original t-J language quasiparticles
with spin-up(down) move in sublattice A(B), the interaction term can
also be  written as a spin-spin interaction. To describe the subspace of zero
spin
of the original t-J model, half the quasiparticles must be
in each sublattice.
This Hamiltonian has been
deduced based on strong AF correlations, and it has a vH singularity in the
noninteracting
DOS, thus we will refer to it as the ``antiferromagnetic-van Hove'' (AFVH)
model.
Pictorially, it is
shown in Fig.2b: quasiparticles move within the A or B sublattices
and interact when they share a link.
Note that in constructing the low energy AFVH Hamiltonian,
retardation effects have been neglected. It will be shown below that
this model leads to d-wave superconductivity, which is mainly induced by
the real space potential
among carriers.\cite{bulut2} Thus, we do not believe that retardation effects
will
change the conclusions of this paper. The good results described below make
this argument self-consistent.

The AFVH model could be studied using powerful
computational techniques, but their
implementation is non-trivial.\cite{computer}  Here, the
analysis of the AFVH Hamiltonian will proceed with the
standard BCS formalism.
Since ${\rm |V|/W \sim 0.2}$, where ${\rm W}$ is the bandwidth of the
quasiparticles, the gap equation should produce a reliable
estimation of the critical temperature since we are effectively
exploring the ``weak'' coupling regime of the AFVH model.
Solving the gap equation on $200 \times 200$ grids,
we observed that the free
energy is minimized using a ${\rm
d_{x^2-y^2}}$ order parameter. After straightforward
algebra it can be shown that the gap equation for the
thermodynamical properties in this channel is
\begin{equation}
{\rm {{1}\over{0.6J}} = {{1}\over{2N}} \sum_{\bf k}
{ {cosk_x (cosk_x - cosk_y) tanh(E_{\bf k}/2T) }\over{E_{\bf k}} } },
\end{equation}
where ${\rm E_{\bf k}}=\sqrt{ ( \epsilon_{\bf k} - \mu )^2 + \Delta^2_{\bf
k}}$, $\Delta_{\bf k} = {{\Delta_0}\over{2}} {\rm (cosk_x - cosk_y)}$,
$\Delta_0$ is the parameter to be
obtained self-consistently, ${\rm N}$ is the number of sites,
and ${\rm T}$ is the temperature.
Note that the T-dependence of
$\mu$ cannot be neglected.
In Fig.2c, ${\rm T_c}$ against the hole density is shown.\cite{com4} Two
features need to be
remarked: i) an optimal doping exists at which ${\rm T_c}$ is maximized
which is a direct consequence of the presence of a large peak in the DOS
of the quasiparticles; ii) the optimal doping (15\%) and
optimal ${\rm T_c}$ of about 100K are in excellent agreement with
the cuprates phenomenology.\cite{pao} Although in the effective AFVH
Hamiltonian the
natural scale of the problem is ${\rm J \sim 1000K}$,
since the ratio
between coupling and bandwidth is small, ${\rm T_c}$ is further reduced
in the weak coupling BCS formalism to about 100K.
Note that this quantitative agreement with experiments is
obtained without the need of ad-hoc fitting parameters.
It is also important to remark that the presence of a finite optimal
density of holes $cannot$ be achieved if a simpler $(cosk_x + cosk_y)^2$
dispersion is used. Thus, as noticed in Ref.\cite{flat-theory}, the
small energy difference between ${\bf k} = (\pi/2,\pi/2)$ and $(\pi,0)$
in the hole dispersion
is crucial for the quantitative success of this approach.

 From the gap equation, the ratio ${\rm R(T) = 2}
\Delta_{max}({\rm T})/{\rm k T_c}$  can be calculated,
(for a d-wave condensate at temperature T, $\Delta_{max}({\rm T})$ is defined
as the maximum value of the gap). In Fig.3a, ${\rm R(T)}$ is
shown as a function of temperature, compared with
recent tunneling data.\cite{gap} At ${\rm T=0}$, the AFVH model predicts ${\rm
R(0) =
5.2}$ and the
tunneling experiment gives 6.2. While the agreement is already
encouraging, note that other experiments have reported a smaller
value for ${\rm R(0)}$. For example, ARPES data by Ma et al.\cite{ma}
obtained
${\rm R(0) = 4.6}$, while an average over the pre-1992 literature\cite{batlogg}
suggested
${\rm R(0) = 5 \pm 1}$ supporting the results of the
AFVH model.
At ${\rm T_c}$
we have also calculated the ratio ${\rm R_2} = \Delta {\rm C / C_{n}}$, where
$\Delta {\rm C = C_{s} - C_{n}}$
is the difference between the specific heat of the normal state ${\rm C_{n}}$
(calculated by turning off the interaction ${\rm |V| }$),
and the superconducting state ${\rm C_{s}}$. The result is shown in Fig.3b.
At ${\rm T_c}$, ${\rm R_2 \approx 4.2}$ which is again in good agreement with
YBCO experiments reported
by Phillips et al., and Loram and Mirza\cite{phillips}
which found ${\rm R_2 = 4.8}$ and ${\rm 4.1}$, respectively.
It is also gratifying that not only dimensionless ratios but also
absolute values are in reasonable agreement with experiments. For
example, in Fig.4a $\Delta {\rm C/ T_c}$ is shown as a function of
density compared with results for YBCO. Although the relation between
hole carriers and the amount of oxygen in this material is unknown,
the general trends between theory and experiment
are similar, as was observed in other vH calculations.\cite{vh}

We have also verified that an important feature of previous vH scenarios
also exists in our model, i.e. a quasiparticle lifetime linear with
frequency at the optimal doping. To carry out the calculation we
switched off the attractive term,
introduce particle-particle repulsion within the $same$ sublattice at
distance one,\cite{com5}
and carry out the standard one bubble approximation to get the imaginary
part of the self-energy. The result
shows a linear behavior at the optimal doping (Fig.4b).
Finally, let us address the stability of the van Hove singularity
against effects not explicitly considered in the Hamiltonian.
In some van Hove theories
the singularity in the DOS is rapidly removed by inhomogeneities
or 3D effects.
For example, the DOS of a 2D tight-binding model
with dispersion $\epsilon_{\bf k} = {\rm -2t (cosk_x + cosk_y)}$ is plotted
in Fig.4c with an energy-independent broadening $\Gamma = 0.02 eV$  using
${\rm N(\omega) = {{2}\over{\pi N}} \sum_{\bf k} {{\Gamma}\over{   (\omega -
\epsilon_{\bf k} )^2 + \Gamma^2}} }$, and ${\rm t=0.4eV}$. Fig.4c shows that
the singularity is unobservable with this broadening.
However, for the dispersion of the AFVH model the situation is different since
the
carriers bandwidth is narrow, with a scale setup by J, and
there is a large accumulation of weight in the DOS at low energy. In
other words, the number of states in the vicinity of the vH singularity is of
the order of the total number of states in the full band. For
these reasons the influence of disorder on the DOS is milder in the AFVH
model, where a large peak can still be seen in Fig.4c even
with a broadening $\Gamma = 0.02 eV$.

In this paper, simple and $quantitative$ ideas
for high-Tc superconductivity have been discussed
combining for the first time two apparently different proposals
for the cuprates i.e. the antiferromagnetic and van Hove scenarios.
We have claimed that these approaches are actually deeply related
with $both$ the pairing mechanism and the vH singularity in the DOS of the
high Tc materials caused by antiferromagnetic correlations. This
theory explains in an economical way the
$d_{x^2 - y^2}$ superconducting state apparently observed in
several experiments, and goes beyond previous work showing that the
critical temperature is maximized at a
particular optimal doping where the quasiparticle lifetime is linear with
energy. An excellent quantitative agreement between theory and experiments is
reported for the critical temperature, optimal doping, superconducting
gap and specific heat.

\medskip
We thank D. S. Dessau, K. Gofron, D. W. Hess, M.
Horbach, J. Kim, D. M. King, J. Ma, R. S. Markiewicz, M. Onellion, J.
Osterwalder,
J. Riera, J. R. Schrieffer, Z.-X. Shen, and B. O. Wells,  for useful
conversations
and suggestions.
E. D. and A. M. are supported by the Office of Naval Research under
grant ONR N00014-93-0495. Part of this work was supported by the
National High Magnetic Field Laboratory, Tallahassee, FL.
\medskip

\vfil\eject

%
% Figure Captions
%

{\bf Figure Captions}

\begin{enumerate}

\item
(a) Quasiparticle dispersion of the 2D t-J model (Eq.(2))
using ${\rm J = 0.125 eV}$. The electron notation is
used, thus all levels below ${\rm E_F}$ are occupied.
Thick lines are the position of the quasiparticles that should be
easily observed in ARPES experiments at a finite hole concentration. The
dashed line represents the ``shadow dispersion'', caused by the remnant
AF at finite density. The calculation was done with a Monte Carlo
technique (Ref.\cite{flat-theory});
(b) Experimental results for Bi2212 (Ref.\cite{aebi})
The ${\rm {\bar M}}$(${\rm Y}$) point corresponds to ${\bf k} =
(\pi,0)$ ($(\pi,\pi)$)
in the 2D square lattice language. The thick line is a strong experimental
signal, while the dashed line is weak;
(c) Direct comparison between experiments and theory: the solid line is
our dispersion Eq.(2),
while the experimental results are from
Ref.\cite{campuzano}.

\item
(a) Spin-spin correlation $C({\bf r})=\langle S^z_{\bf i} S^z_{\bf i+r}
\rangle (-1)^{|{\bf r}|}$ vs distance for the 2D Hubbard model
calculated using Quantum Monte Carlo at
${\rm U/t=8}$ on an $8 \times 8$ cluster. The
$electronic$ densities starting from above are
1.0, 0.90, 0.83, and 0.74. The temperature is ${\rm T = 0.1 eV}$,
$\Delta \tau = 0.0625$, and the error bars are typically about 0.02;
(b) a pictorial representation of the AFVH
model (see text); (c) Critical
temperature ${\rm T_c}$ of the AFVH model as a function of hole
density ${\rm x}$ ($=1-\langle n \rangle)$. The superconducting state is
d-wave.

\item
(a) $2\Delta_{max}(T)/{\rm kT_c}$ against ${\rm T/T_c}$. The solid line
corresponds to the AFVH model at the optimal doping.
The open squares
are tunneling data
for ${\rm Bi_2 Sr_2 Ca Cu_2 O_8}$\cite{gap}.
The full triangle at T=0 corresponds to ARPES
data\cite{ma} while the full square
is a summary of experimental data.\cite{batlogg}
The dotted line is the BCS prediction (i.e. attractive Hubbard model at
half-filling and weak coupling);
(b) ${\rm R_2}$ as defined in the text vs ${\rm T/T_c}$
for the AFVH model  at the
optimal doping. The dot corresponds to experimental results for
YBCO.\cite{phillips}. The BCS result is shown.

\item
(a) ${\rm \Delta C / T_c}$ vs hole density ${\rm x}$ using the AFVH model.
Open squares correspond to experimental data\cite{daum}
for ${\rm Y Ba_2 Cu_3 O_{7-\delta}}$, using the convention that
the result at the optimal $\delta$ ($\sim 0.02$) is plotted at
our optimal doping ${\rm x \sim 0.15}$, and that $\delta \approx {\rm 0.02
+  (0.15-x)}$ in the rest of the data.
The vertical axis is in ${\rm mJ / K^2 / mol}$ units for the
experiment and ${\rm mJ/ K^2 / site}$ for the theory;
(b) Quasiparticle lifetime vs.
energy, $\omega$, at T=0. The solid
line corresponds to the chemical potential, $\mu$, at the
saddle-point, while for the dashed line $\mu$ is above the saddle-point.
Vertical units are arbitrary; (c) DOS
corresponding to the AVFH model (I), and a tight-binding model
(II), using $\Gamma = 0.02eV$.

\end{enumerate}


\begin{references}
\bibitem{com4} We checked that
Hartree-Fock corrections are small, and do not alter the features
of the dispersion Eq.(2).
\bibitem{com5} This interaction does not alter the d-wave
condensate, and it stabilizes it against
phase separation effects.
\end{references}
\end{document}